\documentclass[journal]{IEEEtran}
\ifCLASSINFOpdf
\else
   \usepackage[dvips]{graphicx}
\fi
\usepackage{url}
\usepackage{graphicx,stfloats}
\usepackage[caption=false,subrefformat=parens]{subfig}
\usepackage{amsmath,bm}
\usepackage{booktabs}
\usepackage[style=ieee]{biblatex}
\newcommand{\e}{\mathrm{e}}
\newcommand{\ba}{\mathbf{a}}
\newcommand{\x}{\mathbf{x}}
\newcommand{\w}{\mathbf{w}}
\newcommand{\jj}{\mathcal{J}}
\newcommand{\rr}{\mathbf{r}}
\newcommand{\vv}{\mathbf{v}}
\newcommand{\R}{\mathbf{R}}
\newcommand{\K}{\mathbf{K}}
\newcommand{\trace}{\mathrm{tr}}

\newcommand{\ra}[1]{\renewcommand{\arraystretch}{#1}}

\graphicspath{{Figures/}}

\begin{document}

\title{An Adaptive All-Pass Filter for Time-Varying Delay Estimation}

\author{Beth Jelfs, Shuai Sun, Kamran Ghorbani, and Christopher Gilliam
\thanks{The authors are with the School of Engineering, RMIT University, Melbourne, Australia (e-mail: \{beth.jelfs, shuai.sun2, kamran.ghorbani\}@rmit.edu.au, dr.christopher.gilliam@ieee.org).}}

\maketitle

\begin{abstract}
The focus of this paper is the estimation of a delay between two signals. Such a problem is common in signal processing and particularly challenging when the delay is non-stationary in nature. Our proposed solution is based on an all-pass filter framework comprising of two elements: a time delay is equivalent to all-pass filtering and an all-pass filter can be represented in terms of a ratio of a finite impulse response (FIR) filter and its time reversal. Using these elements, we propose an adaptive filtering algorithm with an LMS style update that estimates the FIR filter coefficients and the time delay. Specifically, at each time step, the algorithm updates the filter coefficients based on a gradient descent update and then extracts an estimate of the time delay from the filter. We validate our algorithm on synthetic data demonstrating that it is both accurate and capable of tracking time-varying delays.
\end{abstract}


\IEEEpeerreviewmaketitle
\section{Introduction}
\label{sec:Intro}
\IEEEPARstart{T}{he} development of algorithms for the estimation of a time delay between two or more spatially separated sensors has a long history~\cite{Knapp1976,Etter1981,Carter1987}. However, recent developments in low cost sensors and the growing ubiquity to deploy large numbers of sensors mean the need for accurate delay estimation techniques is as important as ever~\cite{Wang2016,Sun2016a}. In particular, the estimation of time-varying delays (TVD) occurs in many scenarios involving speed or time of flight measurements. 

A variety of approaches to delay estimation exist, but one set of methods that have been particularly successful are based on adaptive filters~\cite{So1994,Lin1998,Wang2017}. The popularity of the adaptive filter approaches is down to two key attributes. One, they do not require any knowledge of the signal statistics or spectrum \textit{a priori}, requiring only that the source signal is bandlimited and stationary. Two, they are capable of tracking a delay in potentially nonstationary environments, for instance when the sensors move causing the delay between them to change. 

The primary form of an adaptive filter for delay estimation assumes the filter coefficients are samples of a sinc function~\cite{So2001}. Coefficients are then updated using a stochastic gradient descent algorithm, similar to the least mean square (LMS) update~\cite{Widrow1985}. To calculate the delay from the filter coefficients these algorithms often rely on either interpolation~\cite{Feintuch1981} or the use of a lookup table~\cite{So1994}. As such, their accuracy depends on the precision of these methods and can be sensitive to noise. Hence, in~\cite{Chern1994} the direct delay estimation (DDE) technique was proposed as an alternative, overcoming the need for interpolation in the calculation of the delay.

Utilising the assumption that the optimal solution for the adaptation of the filter coefficients is a sinc interpolation filter, and hence all-pass, in~\cite{Sun1999} the authors present a DDE algorithm based on an all-pass constraint. The all-pass constraint is both more general and more flexible than the previous constraints and the algorithm was shown to outperform other methods, particularly for low signal-to-noise ratios (SNRs). However, despite the relative flexibility of this method, in practice it estimates a truncated version of a sinc interpolation filter, which can potentially affect the accuracy of the delay estimate. Specifically, the truncated filter has a finite impulse response (FIR) and a FIR filter can only be truly all-pass if it has only one non-zero coefficient. This situation occurs only when the delay is integer in nature. For non-integer delays to obtain a true all-pass filter requires the use of an infinite impulse response (IIR). 

In~\cite{Gilliam2018} an all-pass framework was introduced for image alignment. This framework leveraged the fact that any all-pass filter can be defined as a ratio of a forward filter and its time reversal, backward, filter to create an FIR representation of the IIR. Previously, we have shown this framework can be used to formulate an effective TVD estimator using short local windows of data~\cite{ICASSP2018}. However, in order to allow the estimation of delays on different scales this framework requires multiple filters to be combined sequentially, with post-processing at each stage. Hence, requiring the entirety of the input signals to be recorded before an estimate of the delay can be obtained. 

In this paper we propose a new linear predictor which allows us to combine the all-pass framework for delay estimation with an adaptive filter for TVD estimation. The proposed implementation updates the all-pass filter at each time step using a normalised LMS (NLMS) style algorithm. This solution provides a normalised adaptive all-pass (NAAP) filter capable of estimating and tracking TVDs without the need for post-processing. Simulations demonstrate the efficacy of our implementation with the results showing the proposed algorithm is accurate and more versatile than alternative methods.

\section{Delay Estimation Using All-Pass Filters}
\label{sec:APFramework}
Before introducing the proposed NAAP filter we first outline a delay estimation framework based on all-pass filters for a constant delay. This framework is the underlying principle of the algorithm presented in~\cite{Gilliam2018,ICASSP2018} and forms the basis of our approach. The framework revolves around the equivalence between a constant delay and filtering with an all-pass filter. From the Fourier shift theorem: if we have a signal, $x$, then delaying $x$ by a constant, $\tau$, is equivalent to multiplication by a complex exponential in the frequency domain. This operation can be interpreted as filtering by a filter, $h$, with the frequency response
\begin{equation}
	H(\omega) = \e^{-j\tau\omega},
	\label{eq:Fourier_Shift}
\end{equation}
where $H$ is the Fourier transform of $h$ and $\omega$ denotes the frequency coordinate. This filter has a real-valued impulse response (i.e. $H(\omega) = H^{\ast}(-\omega)$); is all-pass in nature (i.e. $|H(\omega)| = 1$); and, importantly, its phase depends on the time delay. Hence, obtaining the time delay can be achieved by estimating the filter $h$ and then extracting an estimate of $\tau$. 

If we assume ideal sampling, then the 2$\pi$-periodic frequency response of any digital all-pass filter can be represented as~\cite{Regalia1998}
\begin{equation}
	H(\omega) = \frac{P\left(\e^{j\omega}\right)}{P\left(\e^{-j\omega}\right)},
	\label{eq:Ratio_representation}
\end{equation}
where $P\left(\e^{j\omega}\right)$ is the frequency response of the forward FIR filter $p$ and $P\left(\e^{-j\omega}\right)$ the corresponding backward version of same filter. This property allows the all-pass filtering operation performed by $h$ to be written linearly in terms of $p$ as follows:
\begin{equation}
	x(n-\tau) = h(n)\ast x(n) \ \Longleftrightarrow \ p(-n)\ast x(n-\tau) = p(n) \ast x(n),
	\label{eq:AP_equation}
\end{equation}
where $\ast$ is the convolution operator.

Assuming a FIR filter with finite support $k\in[0,K]$ then the filter response $p(k)$ can be described by the coefficients $a_k$ such that
\begin{equation}
    p(k) = \begin{cases}
            a_k, & 0\leq k \leq K \\
            0, & \text{otherwise},
        \end{cases}
\end{equation}
giving
\begin{equation*}
    p(n) \ast x(n) = \sum_{k=0}^K a_kx(n-k). 
\end{equation*}
We can then rewrite the convolution in~\eqref{eq:AP_equation} as
\begin{equation}
    \sum_{k=0}^K a_kx(n+k-\tau) = \sum_{k=0}^K a_kx(n-k).
	\label{eq:filter_relationship}
\end{equation}
The problem then becomes to estimate the filter coefficients $\{a_k\}_{k=0,...,K}$, where $K$ also defines the maximum delay which can be estimated by the filter. Having estimated the filter coefficients we can directly estimate the delay by evaluating $\mathrm{d}H(\omega)/\mathrm{d}\omega$ at $\omega=0$ and equating~\eqref{eq:Fourier_Shift} to~\eqref{eq:Ratio_representation} giving:
\begin{equation}
	\hat{\tau} = 2\frac{\sum_k ka_k}{\sum_k a_k}.
	\label{eq:delay}
\end{equation}

\subsection{Proposed Linear Predictor}
\label{ssec:proposed}
Taking advantage of~\eqref{eq:delay} we now introduce a linear predictor which utilises the relationship~\eqref{eq:filter_relationship} to provide an estimate of the delay. If we consider two spatially separated sensors which both receive signal $x$ but with a delay between them, then at sample time $n$, sensor 1 receives $x(n)$ and sensor 2 receives $x(n-\tau)$. This is equivalent to assuming $a_0=1$ and we can rearrange~\eqref{eq:filter_relationship} to obtain
\begin{align}
	x(n\!-\!\tau) - x(n) &= \sum_{k=1}^K a_kx(n-k) - \sum_{k=1}^K a_kx(n+k-\tau) \nonumber\\
	&= \x_-^T(n)\ba - \x_+^T(n-\tau)\ba,
	\label{eq:current_samples}
\end{align}
where $\ba = \left[a_1,\ldots,a_K\right]^T$, $\x_-(n) = \big[x(n\!-\!1),\ldots,x(n\!-\!K)\big]^T$ is the backward vector of sensor 1 and $\x_+(n-\tau) = \big[x(n\!+\!1\!-\!\tau),\ldots,x(n\!+\!K\!-\!\tau)\big]^T$ is the forward vector of sensor 2. This allows us to express the current samples obtained at each of the sensors as a linear predictor of the samples from the other sensor such that
\begin{align}
	x(n) &= \x_+^T(n-\tau)\ba, \label{eq:x_filter}\\
	x(n-\tau) &= \x_-^T(n)\ba. \label{eq:xtau_filter}
\end{align}

To illustrate this relationship, take the filter with $K=3$:
\begin{align*}
	x(n) =&\ \x_+^T(n-\tau)\ba \\
	=&\ a_1x(n\!+\!1\!-\!\tau) + a_2x(n\!+\!2\!-\!\tau) + a_3x(n\!+\!3\!-\!\tau), \\
	x(n\!-\!\tau) =&\ \x_-^T(n)\ba \\
	=&\ a_1x(n-1) + a_2x(n-2) + a_3x(n-3),
\end{align*}
for a simple integer delay, such as $\tau=2$ this gives
\begin{align*}
	x(n) =&\ a_1x(n-1) + a_2x(n) + a_3x(n+1), \\
	x(n-2) =&\  a_1x(n-1) + a_2x(n-2) + a_3x(n-3),
\end{align*}
and hence $\ba = \begin{bmatrix} 0 & 1 & 0 \end{bmatrix}$ which substituting into~\eqref{eq:delay}, noting $a_0=1$, gives $\hat{\tau} = 2\big(0\times1 + 1\times0 + 2\times1 + 3\times0\big)/\big(1 + 0 + 1 + 0\big) = 2$. In the following, we build upon this framework to develop an adaptive filtering algorithm capable of estimating time-varying delays.

\section{Adaptive All-Pass Filter}
\label{sec:AAPFilter}
To introduce our NAAP algorithm we assume, without loss of generality, that the desired response can be represented as the difference between the noisy outputs of the optimum filter such that~\cite{Douglas1995}
\begin{equation}
	d(n) = \x_-^T(n)\ba + \eta_1(n) - \x_+^T(n-\tau)\ba - \eta_2(n), 
	\label{eq:desired}
\end{equation}
where $\eta_1(n)$ and $\eta_2(n)$ are zero mean i.i.d. noise sources with variance $\sigma_\eta^2$ and are independent of $x(n)$. We can then define our current estimate of the true filter coefficients $\ba$ as $\w(n) = \left[w_1,w_2,\ldots,w_K\right]^T$ and obtain the filter output
\begin{equation}
	y(n) = \Big[\x_-^T(n) - \x_+^T(n-\tau)\Big]\w(n).
	\label{eq:output}
\end{equation}
Our problem thus becomes one of minimizing the error between the measured samples, $d(n)$ and their estimates obtained from the current filter coefficients $y(n)$
\begin{align}
	e(n) =&\ d(n) - y(n) \nonumber\\
	=&\ \Big[\x_-^T(n) - \x_+^T(n-\tau)\Big]\ba \nonumber\\
	&- \Big[\x_-^T(n) - \x_+^T(n-\tau)\Big]\w(n) + \eta_1(n) - \eta_2(n). 
	\label{eq:error}
\end{align}

To obtain an adaptive all-pass filter we define the update based on the steepest-descent as
\begin{equation}
	\w(n+1) = \w(n) - \mu\nabla\mathcal{J}(n)\big|_{w=w(n)},
	\label{eq:w_cost}
\end{equation}
with learning rate, $\mu$. Using the cost function
\begin{align}
	\jj(n) =& \left|e(n)\right|^2, \label{eq:cost}\\
	\intertext{gives the gradient}
	\nabla\jj(n)\big|_{w=w(n)} =& \frac{\partial\jj(n)}{\partial\w(n)} = 2e(n)\frac{\partial e(n)}{\partial\w(n)} \nonumber\\
	=& -2e(n)\Big[\x_-^T(n) - \x_+^T(n-\tau)\Big], \label{eq:gradient}
\end{align}
which results in the adaptive all-pass update:
\begin{equation}
	\w(n+1) = \w(n) + 2\mu e(n)\Big[\x_-^T(n) - \x_+^T(n-\tau)\Big].
	\label{eq:w_update}
\end{equation}

If we define $\rr(n)$ as the residual difference between $\x_-(n)$ and $\x_+(n-\tau)$ and $\eta(n) = \eta_1(n)-\eta_2(n)$ this leads to the adaptive all-pass filter in precisely the form of the standard LMS algorithm~\cite{Widrow1985} for the input vector $\rr(n)$
\begin{align}
	e(n) =&\ \rr^T(n)\ba - \rr^T(n)\w(n) + \eta(n), \label{eq:error_r}\\
	\w(n+1) =&\ \w(n) + 2\mu e(n)\rr^T(n). \label{eq:w_update_r}
\end{align}
Therefore, following the standard analysis for the LMS algorithm~\cite{Feuer1985,Farhang-Boroujeny2013} we can show (see Appendix~\ref{ssec:Behaviour}) 
the mean square error $\xi(n)$ converges asymptotically to $\xi(\infty) = \xi_{min} = \sigma_\eta^2$ for
\begin{equation}
	0<\mu<\frac{1}{3\trace[\R]}\ \ ,
	\label{eq:bound}
\end{equation}
where $\trace[\cdot]$ is the trace of the matrix. We note that unlike the standard LMS, the input vector here is the residual signal $\rr(n)$ and therefore the correlation is based on the difference between the forward and backward vectors from the two sensors, such that $\R = \big[\x_-(n) - \x_+(n\!-\!\tau)\big]\big[\x_-(n) - \x_+(n\!-\!\tau)\big]^T$.

\subsection{Normalised Adaptive All-Pass Filter}
\label{ssec:NAAP}
For convergence in the mean square the bound on the learning rate is given by~\eqref{eq:bound} where $\trace[\R]$ is the instantaneous energy of the residual signal
\begin{equation*}
    \trace[\R] =\rr^T(n)\rr(n) = \sum_{i=1}^K r^2(n-i),
\end{equation*}
and $r(n-i)=x(n-i)-x(n+i-\tau)$ which leads to the bound on the learning rate
\begin{equation*}
	0<\mu<\frac{1}{3\|\x_-(n)-\x_+(n-\tau)\|_2^2}.
\end{equation*}
Incorporating this into the adaptive all-pass filter gives the normalised adaptive all-pass (NAAP) filter
\begin{align*}
	\w(n+1) = \w(n) + \frac{\rho}{\|\x_-(n)-\x_+(n-\tau)\|_2^2 +\varepsilon}e(n)\rr(n),
\end{align*}
where $0<\rho<1/3$ and $\varepsilon$ is a small positive regularisation constant.

\section{Results}
\label{sec:Results}
\begin{figure}[tb!]
	\centering
	\includegraphics[width=\linewidth]{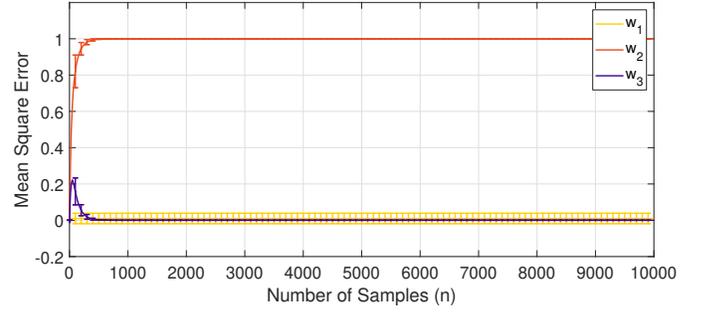}
	\caption{Average evolution of the filter weights for K=3, $\tau=2$. Error bars indicate $\pm2$ standard deviations.}\vspace*{-3mm}
	\label{fig:Weights}
\end{figure}

In this section we evaluate our proposed NAAP filter. To illustrate the NAAP algorithm performs as expected we first analyse the filter weights for the example given in Section~\ref{sec:APFramework}. Figure~\ref{fig:Weights} shows that for the optimum filter $\ba = \begin{bmatrix} 0 & 1 & 0 \end{bmatrix}$, the filter weights converge on average to $w_1 = 0.01 \pm 0.01$, $w_2=1.00 \pm 0.00$ and $w_3 = 0.00 \pm 0.00$ (mean$\pm$SD).

In the following we compare the NAAP to the explicit time delay estimator (ETDE)~\cite{So1994} and the adaptive algorithm proposed by Sun and Douglas in~\cite{Sun1999}, which we term `Sun'. In all of the following simulations $K=7$. The first signal, $x(n)$, was generated by drawing samples from a zero-mean Gaussian distribution then bandlimiting to a bandwidth of $\pi/2$ (normalised frequency). The second signal, $x(n-\tau)$, was obtained via high quality interpolation~\cite{Blu2001} using the ground truth delay signal. Finally, statistical averages were obtained using 100 realisations of the synthetic signals\footnote{Code to reproduce these results is provided in our GitHub repository: \url{https://github.com/beteje/LAP_DelayEstimation}}.

\subsection{Constant Delay}
\label{ssec:ConstDelay}
 
\begin{figure*}[tb!]
	\centering
	\subfloat[]{
	    \includegraphics[width=0.48\linewidth]{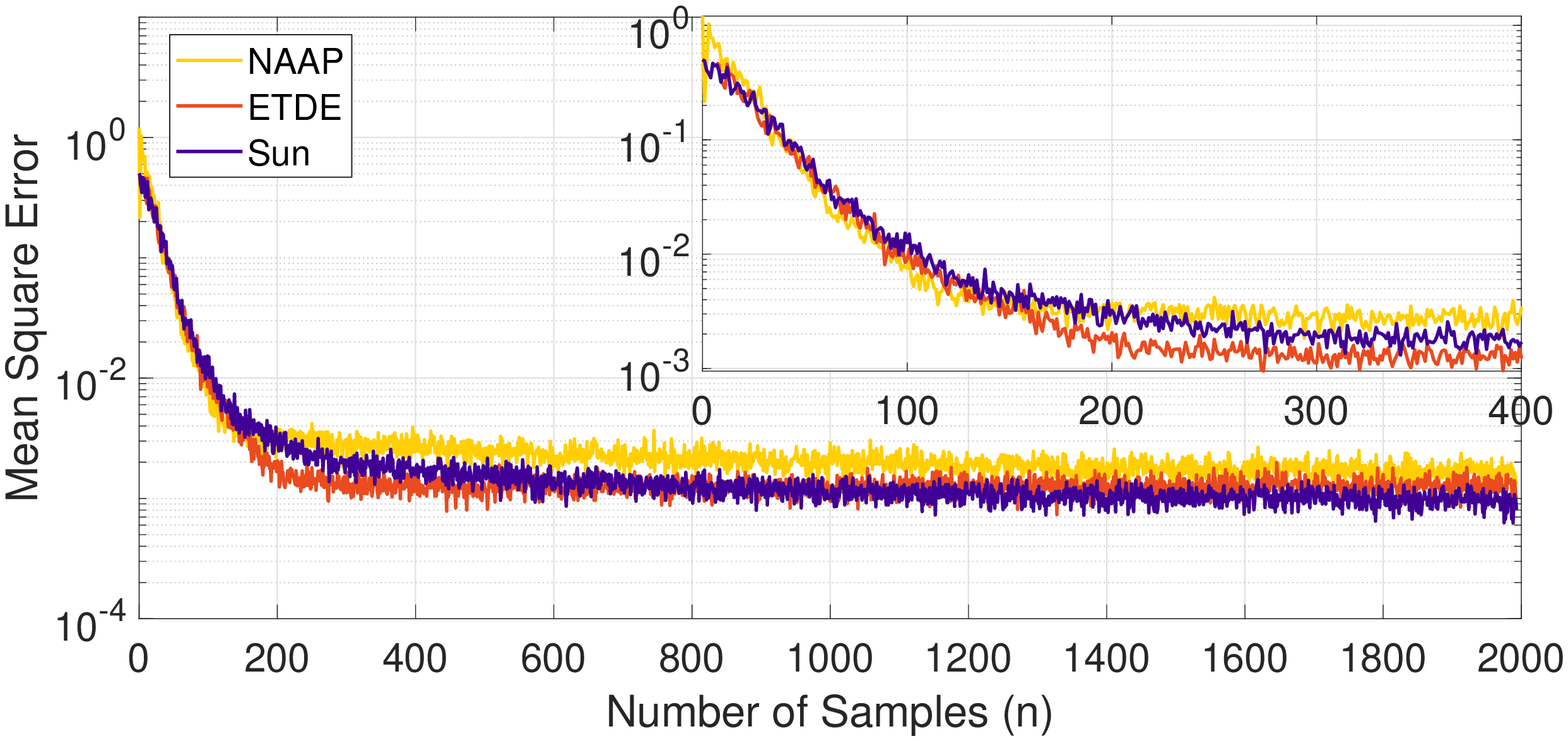}
	    \label{sfig:Convergence}
    }\hfil
    \subfloat[]{
        \includegraphics[width=0.48\linewidth]{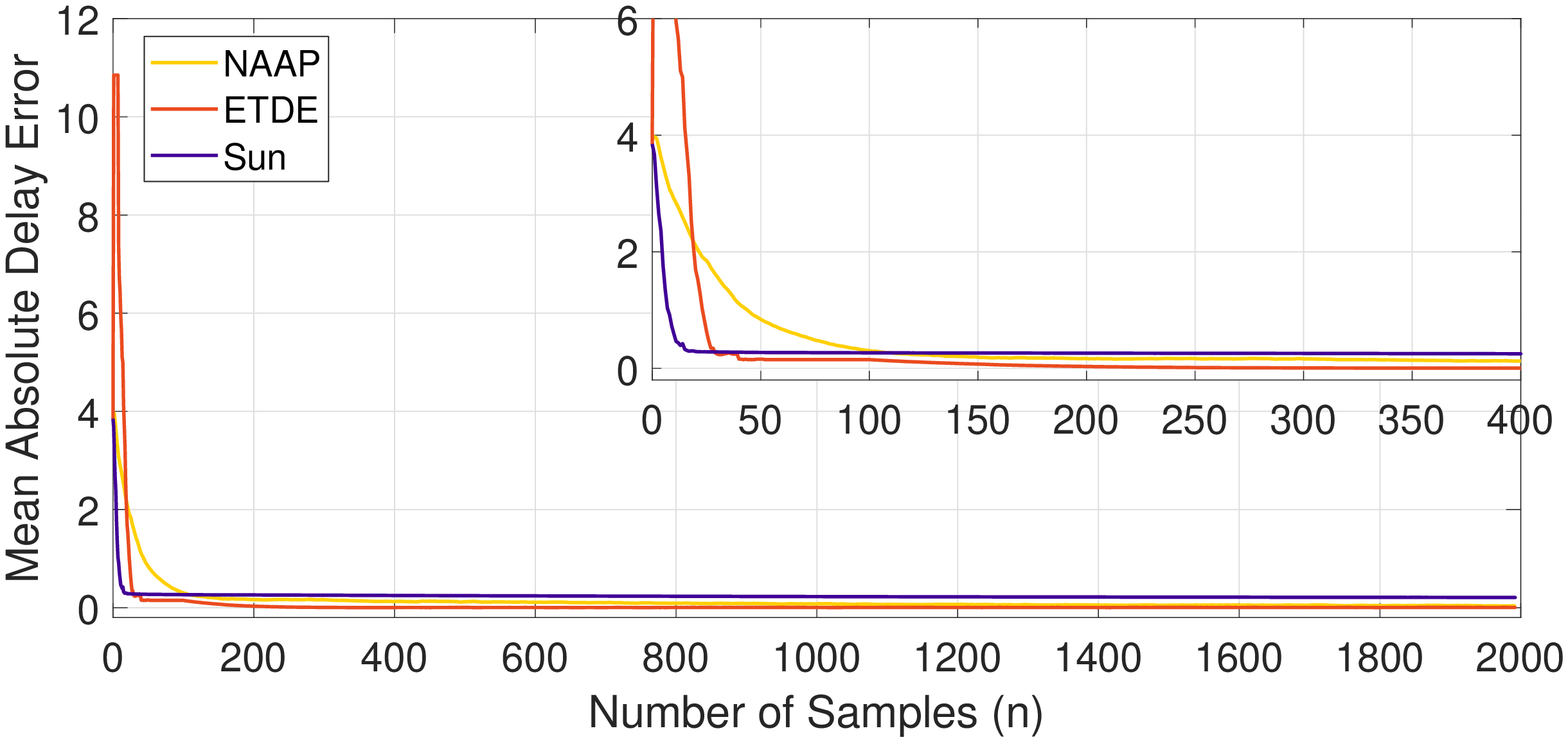}
        \label{sfig:Delay_estimate}
    }
	\caption{(a) Average evolution of the MSE for the NAAP ($\rho=0.08$), the ETDE~\cite{So1994} ($\mu=0.04$) and Sun algorithm~\cite{Sun1999} ($\mu=0.02$). (b) Evolution of the mean absolute delay error, in samples, for the NAAP, ETDE and Sun algorithms. Insets: the first 400 samples.}\vspace*{-6mm}
\end{figure*}

First we analyse the convergence of the NAAP when estimating a constant non-integer delay of $\tau(n) = 5.85$ samples. Figure~\subref*{sfig:Convergence} shows the mean square error (MSE) for the three algorithms, the learning rates were chosen so as to give approximately the same rate of convergence. In terms of the MSE all three algorithms provide similar performance, however, the goal is to obtain the delay between the signals. Obtaining estimates of the delay from the filter weights we can observe differences in the performances of the algorithms.

Although the convergence rates of the filters are the same, Fig.~\subref*{sfig:Delay_estimate} shows that their convergence in the delay error, that is the error between the estimated and true delay, differ. The Sun algorithm achieves the fastest rate of convergence in the delay but a slightly higher steady-state delay error. The NAAP provides a similar steady-state delay error to the ETDE but with the slowest convergence. Our proposed algorithm, however, still converges within approximately 100 samples and as we show in the following section provides greater flexibility.

\begin{figure*}[tb!]
	\centering
	\subfloat[]{
        \includegraphics[width=0.48\linewidth]{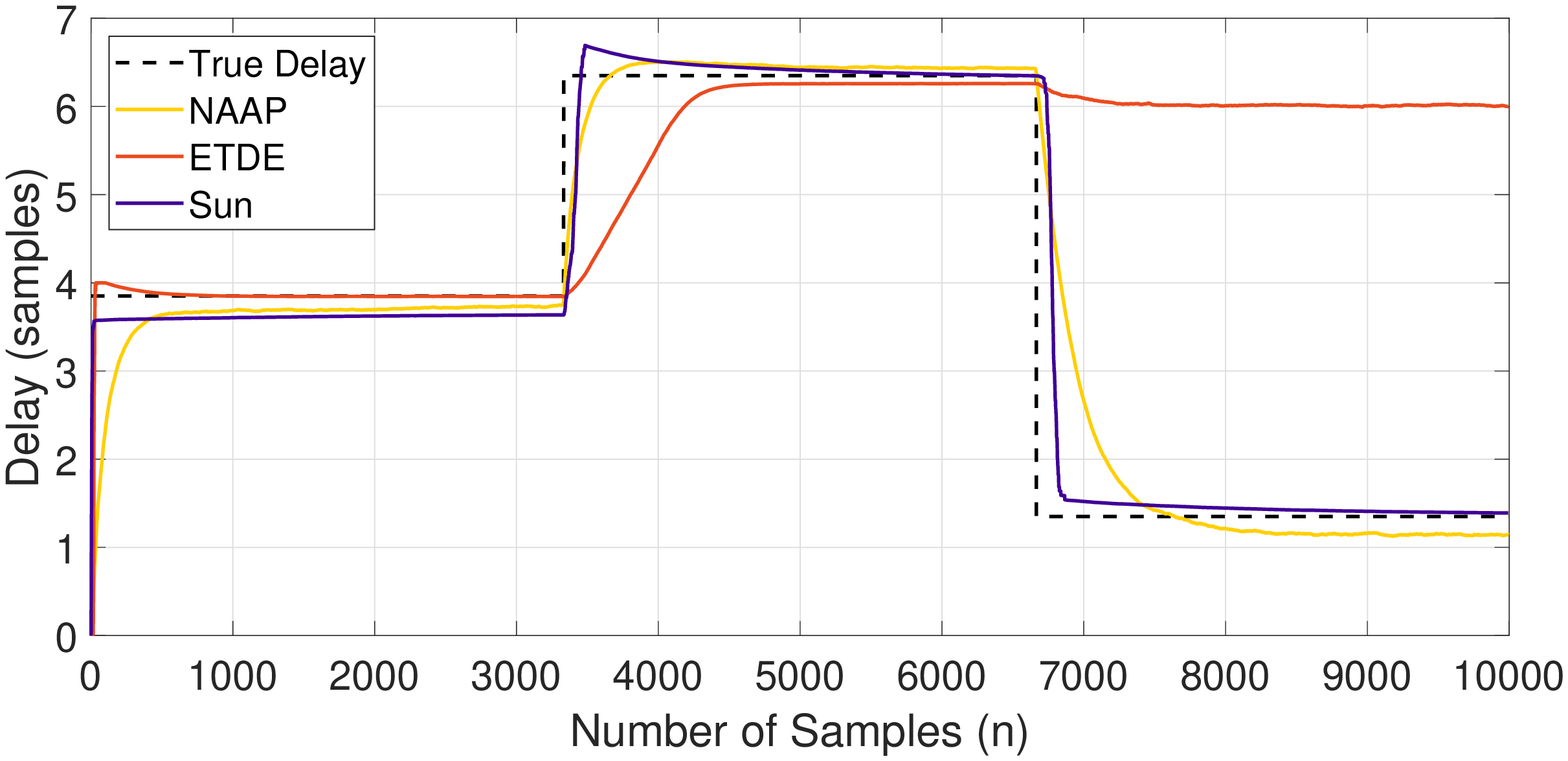}
        \label{sfig:SNR10Tracking}
    }\hfil
    \subfloat[]{
        \includegraphics[width=0.48\linewidth]{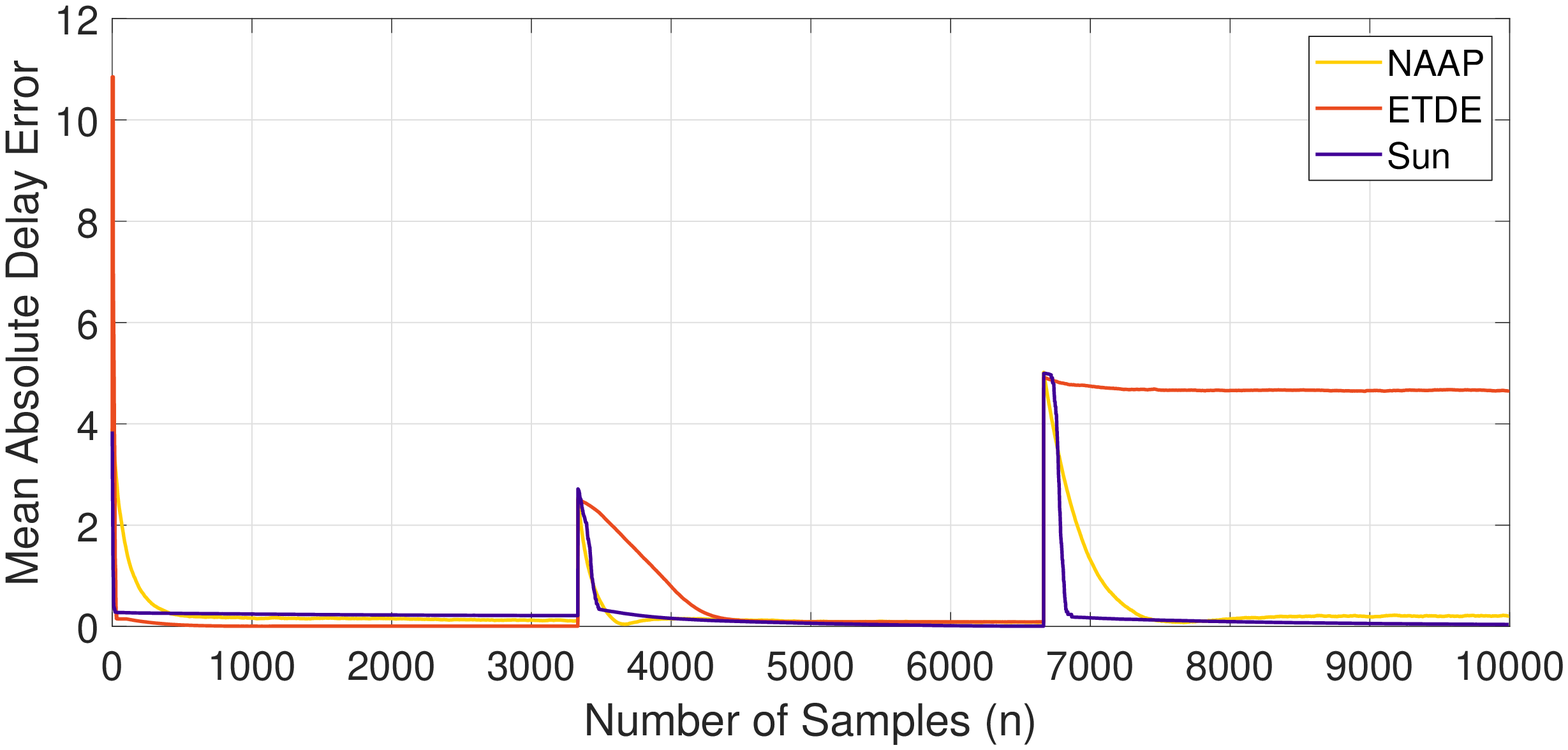}
        \label{sfig:SNR10Error}
    }
	\caption{(a) Average evolution of the delay for the NAAP ($\rho=0.01$), the ETDE ($\mu=0.02$) and Sun algorithm~ ($\mu=0.008$) for the large step change scenario with SNR=20dB. (b) The corresponding evolution of the mean absolute delay error, in samples, for the NAAP, ETDE and Sun algorithms.}
	\label{fig:Tracking}
\end{figure*}

\subsection{Tracking Performance}
\label{ssec:Tracking}
To illustrate the performance of the NAAP algorithm in changing environments, a piecewise constant delay signal with two step changes was used. Two scenarios were implemented both with an initial delay of $\tau=3.85$ samples. The first scenario used `small' step changes ($+0.75$ and $-1.50$ samples). Whereas the second scenario used `large' step changes ($+2.50$ and $-5.00$ samples). For both scenarios SNRs of 5dB, 10dB, 20dB and 30dB were tested. To compare the performance of the algorithms in each of the scenarios the mean absolute delay error was averaged over time to provide a summary statistic and the learning rates chosen so as to minimise this error. The analysis of the different learning rates is provided in Appendix~\ref{ssec:LearningRate}. 

The results given in Table~\ref{tab:SNR} show the ETDE performed well in all noise levels. However, in the large step change scenario the algorithm could not track the largest step change (regardless of the choice of learning rate) as illustrated in Fig.~\ref{fig:Tracking}. This limits application to scenarios where we know the delay will have only small changes over time. In comparison both the NAAP and the Sun algorithm were capable of tracking the delays in both scenarios.

Comparing the results from the NAAP and the Sun algorithm we can see that in the scenarios with large noise values the Sun algorithm outperformed the proposed NAAP algorithm. However, empirically we found the Sun algorithm is extremely sensitive to the choice of learning rate. This was particularly true in the low SNR scenarios where a small change in the learning rate led to the algorithm becoming numerically unstable (see Appendix~\ref{ssec:LearningRate}). 
In comparison the NAAP performed well across all SNRs in both step change scenarios and did not exhibit undue sensitivity to its parameters.

\begin{table}[tb]
	\caption{Average mean absolute delay errors for different SNR values with small step or large step changes.} 
	\label{tab:SNR} 
	\centering 
	\ra{1.2}
	\begin{tabular}{l c@{\hspace{5mm}} c@{\hspace{5mm}} c@{\hspace{5mm}} c} 
		\toprule
		SNR (dB) & 5 & 10 & 20 & 30 \\
		\midrule
		\multicolumn{5}{c}{\textbf{Small Step Change}} \\
		\midrule
		NAAP ($\rho=0.01$) & 0.496 & 0.313 & 0.153 & 0.124 \\
		ETDE ($\mu=0.02$) & 0.112 & 0.074 & 0.052 & 0.047 \\
		Sun ($\mu=0.008$) & 0.249 & 0.235 & 0.235 & 0.234 \\
		\midrule
		\multicolumn{5}{c}{\textbf{Large Step Change}} \\
		\midrule
		NAAP ($\rho=0.01$) & 0.528 & 0.337 & 0.228 & 0.219 \\
		ETDE ($\mu=0.02$) & 1.805 & 1.700 & 1.661 & 1.663 \\
		Sun ($\mu=0.008$) & 0.243 & 0.230 & 0.233 & 0.233 \\
		\bottomrule
	\end{tabular}
\end{table}

\section{Conclusions}
\label{sec:conc}
In this paper we presented an adaptive all-pass filtering algorithm for time-varying delay estimation. Our approach is based on exploiting the fact that any all-pass filter can be represented as a ratio of a forward FIR filter and its time reversal. Using this property, we formulated a LMS style algorithm to estimate the coefficients of the FIR filter, and hence the all-pass filter, at each time step. The time delay is estimated using a direct expression based on the filter coefficients. Simulations demonstrated that our adaptive algorithm converges in the MSE and for time-varying delays provides a more versatile estimate than alternative methods.

\appendix
\subsection{Behaviour of the Adaptive All-Pass Filter}
\label{ssec:Behaviour}
For the adaptive all-pass filter in the form:
\begin{align}
	e(n) =&\ \rr^T(n)\ba - \rr^T(n)\w(n) + \eta(n),\\
	\w(n+1) =&\ \w(n) + 2\mu e(n)\rr^T(n). 
\end{align}
where $\rr = \Big[\x_-(n) - \x_+(n-\tau)\Big]$ and following the analysis for the LMS algorithm~\cite{Feuer1985,Farhang-Boroujeny2013} 
we can substitute in the weight error vector $\vv(n) = \w(n) - \ba$ to give the error as
\begin{equation}
	e(n) = \eta(n) - \rr^T(n)\vv(n).
	\label{eq:error_v}
\end{equation}
Subtracting the optimal weight vector $\ba$, squaring both sides and taking the expectation gives the mean square error
\begin{equation}
	E\Big[e^2(n)\Big] = \sigma_\eta^2 + E\Big[\big(\rr^T(n)\vv(n)\big)^2\Big] - 2E\Big[\eta(n)\rr^T(n)\vv(n)\Big].
	\label{eq:mse}
\end{equation}
Next, employing the standard independence assumptions (namely that the input signal and filter coefficient vectors are zero mean, stationary, jointly normal and with finite moments) the final term in~\eqref{eq:mse} disappears and we can obtain
\begin{align}
	\xi(n) = E\Big[e^2(n)\Big] &= \xi_{min} + \xi_{EMSE}(n) = \sigma_\eta^2 + \trace\Big[\K(n)\R\Big],
	\label{eq:mse_emse}
\end{align}
where $\K(n) = E\big[\vv(n)\vv^T(n)\big]$ is the weight error vector correlation matrix, $\xi_{min}$ is the minimum mean square error defined by the power of the noise $\sigma^2_\eta$, and $\xi_{EMSE}(n)$ is the excess mean square error. The excess means square error is a result of the filter coefficients fluctuating around their optimum values as they begin to converge. Considering the mean square performance in terms of the misadjustment and following the standard approach for the LMS algorithm the misadjustment can be shown to be~\cite{Farhang-Boroujeny2013}
\begin{equation}
	\mathcal{M} = \frac{\xi_{EMSE}}{\xi_{min}} = \mu\frac{\trace[\R]}{1-\mu\trace[\R]}\ \ , 
	\label{eq:m}
\end{equation}
and the mean square error $\xi(n)$ converges asymptotically to $\xi(\infty) = \xi_{min} = \sigma_\eta^2$ for $0<\mu<\frac{1}{3\trace[\R]}$.

\subsection{Learning Rate Comparison}
\label{ssec:LearningRate}
\begin{figure*}[htb]
	\centering
	\vspace*{-5mm}\subfloat[]{
	    \includegraphics[width=0.32\linewidth]{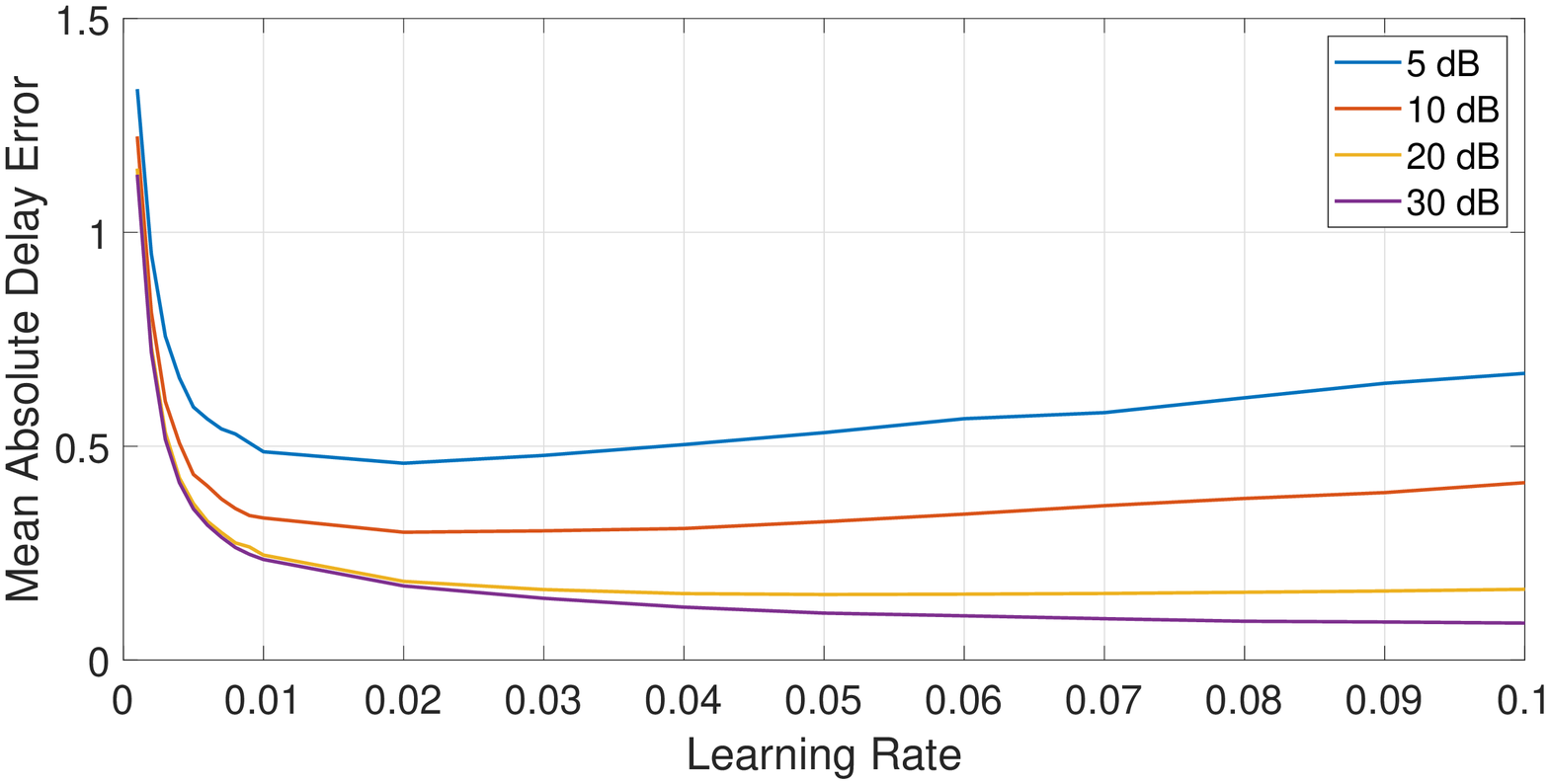}
    }
    \subfloat[]{
	    \includegraphics[width=0.32\linewidth]{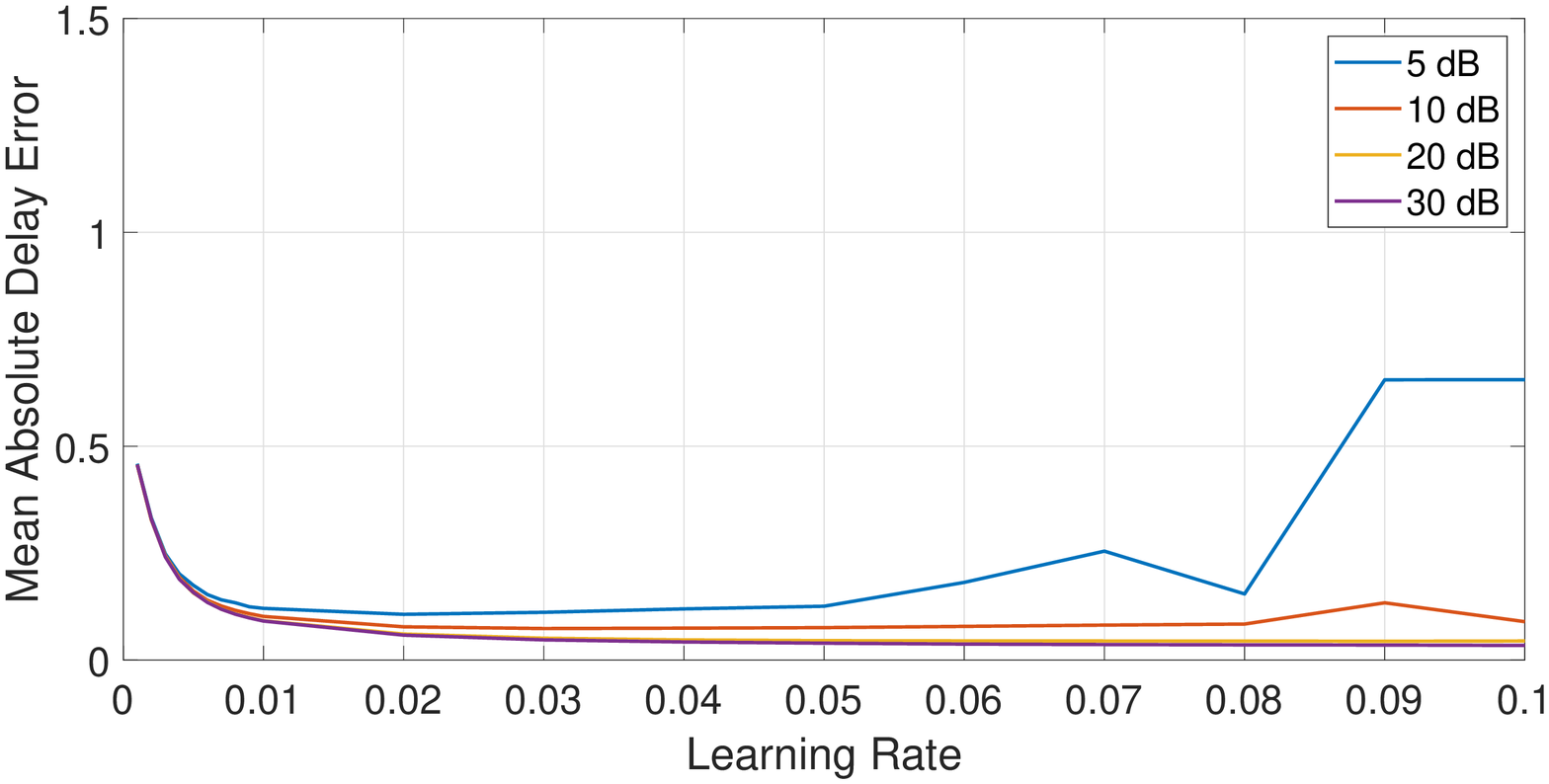}
    }
    \subfloat[]{
	    \includegraphics[width=0.32\linewidth]{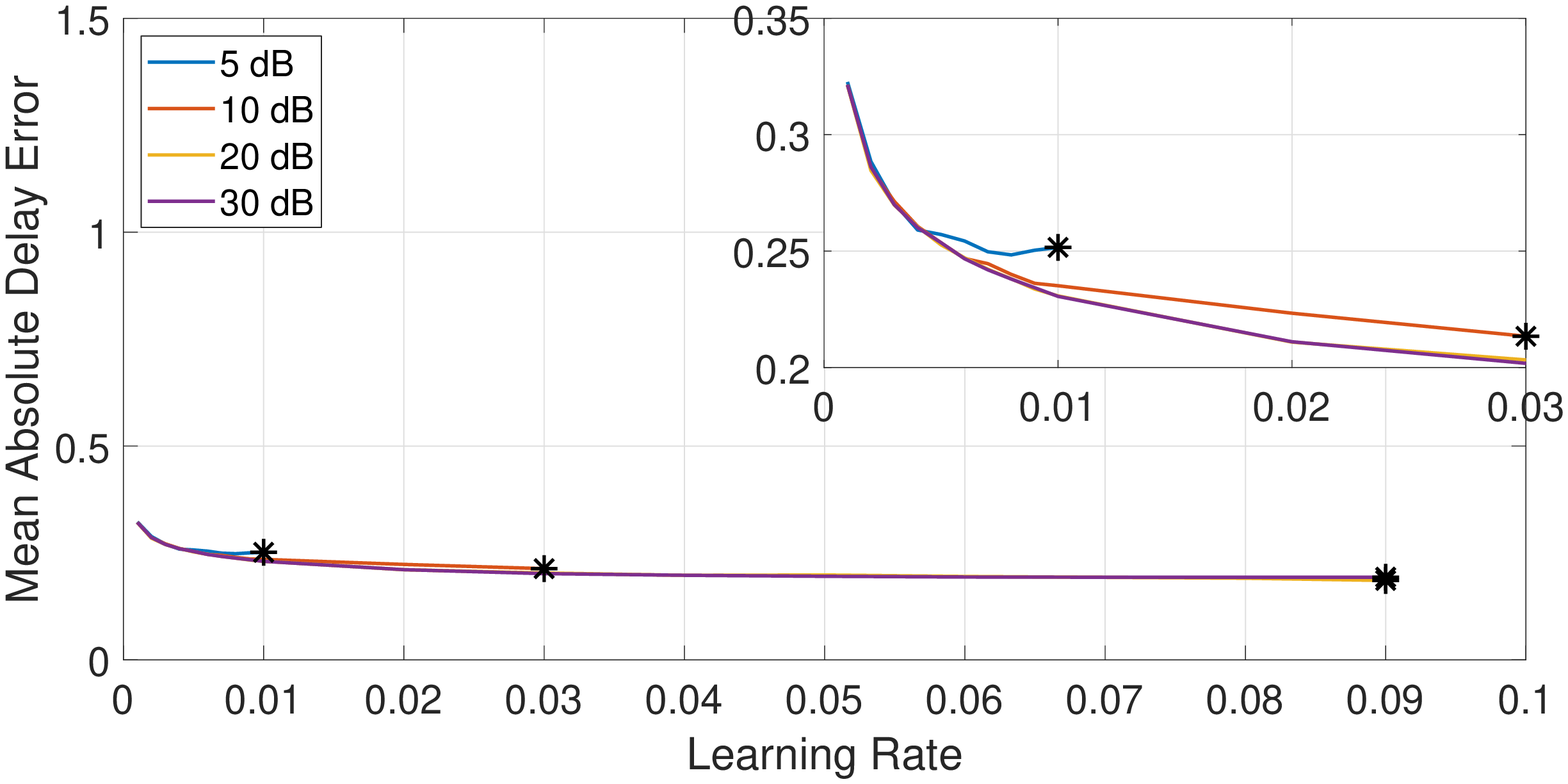}
    }
	\caption{Average mean absolute delay error for different learning rates in the small step change scenario (a) The NAAP. (b) The ETDE (c) The Sun algorithm. The marker $\mathbf{\ast}$ indicates the point at which the algorithm becomes numerically unstable.}
	\label{fig:smallStep}
\end{figure*}

\begin{figure*}[htb]
	\centering
	\vspace*{-5mm}\subfloat[]{
	    \includegraphics[width=0.32\linewidth]{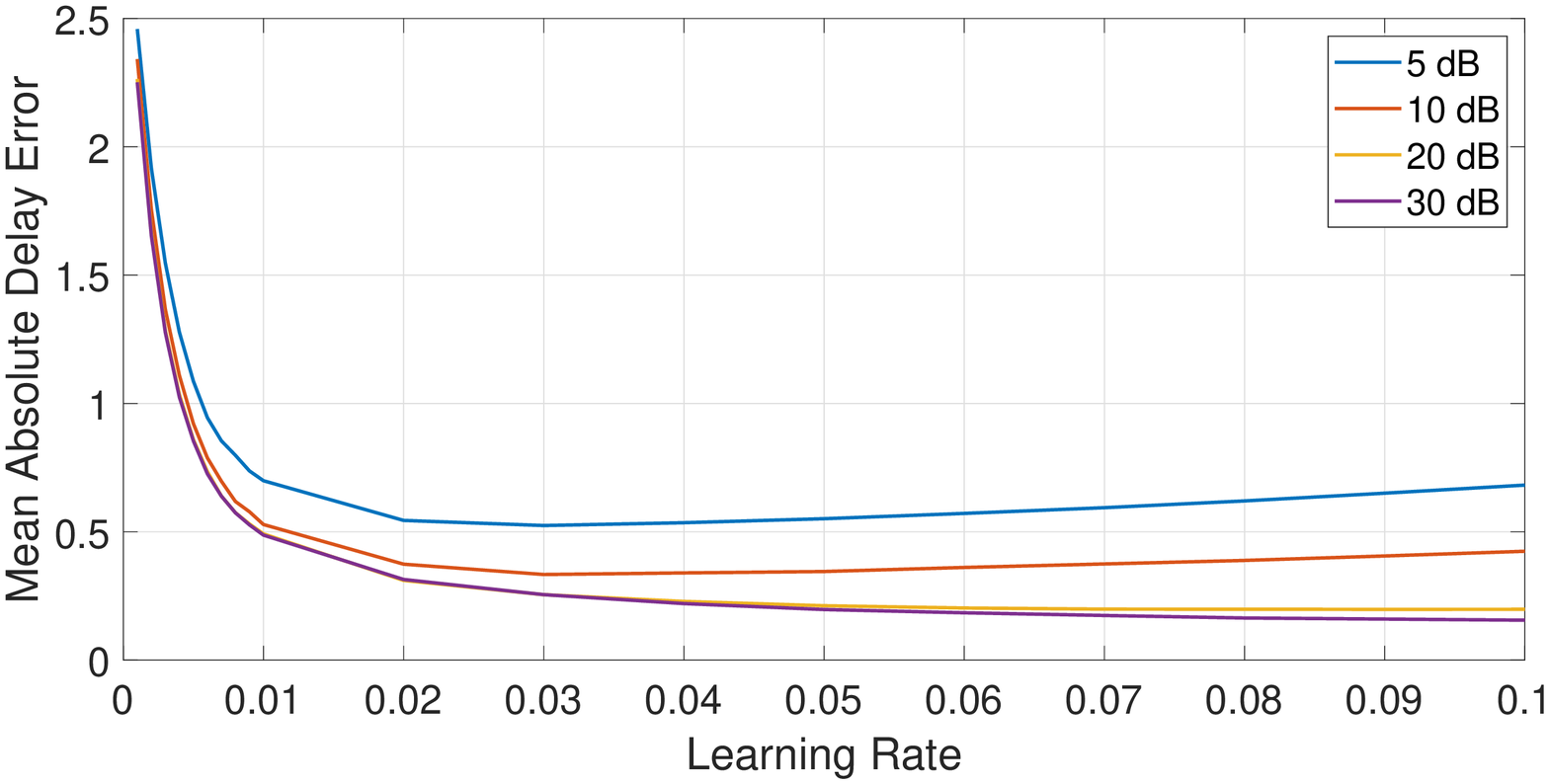}
    }
    \subfloat[]{
	    \includegraphics[width=0.32\linewidth]{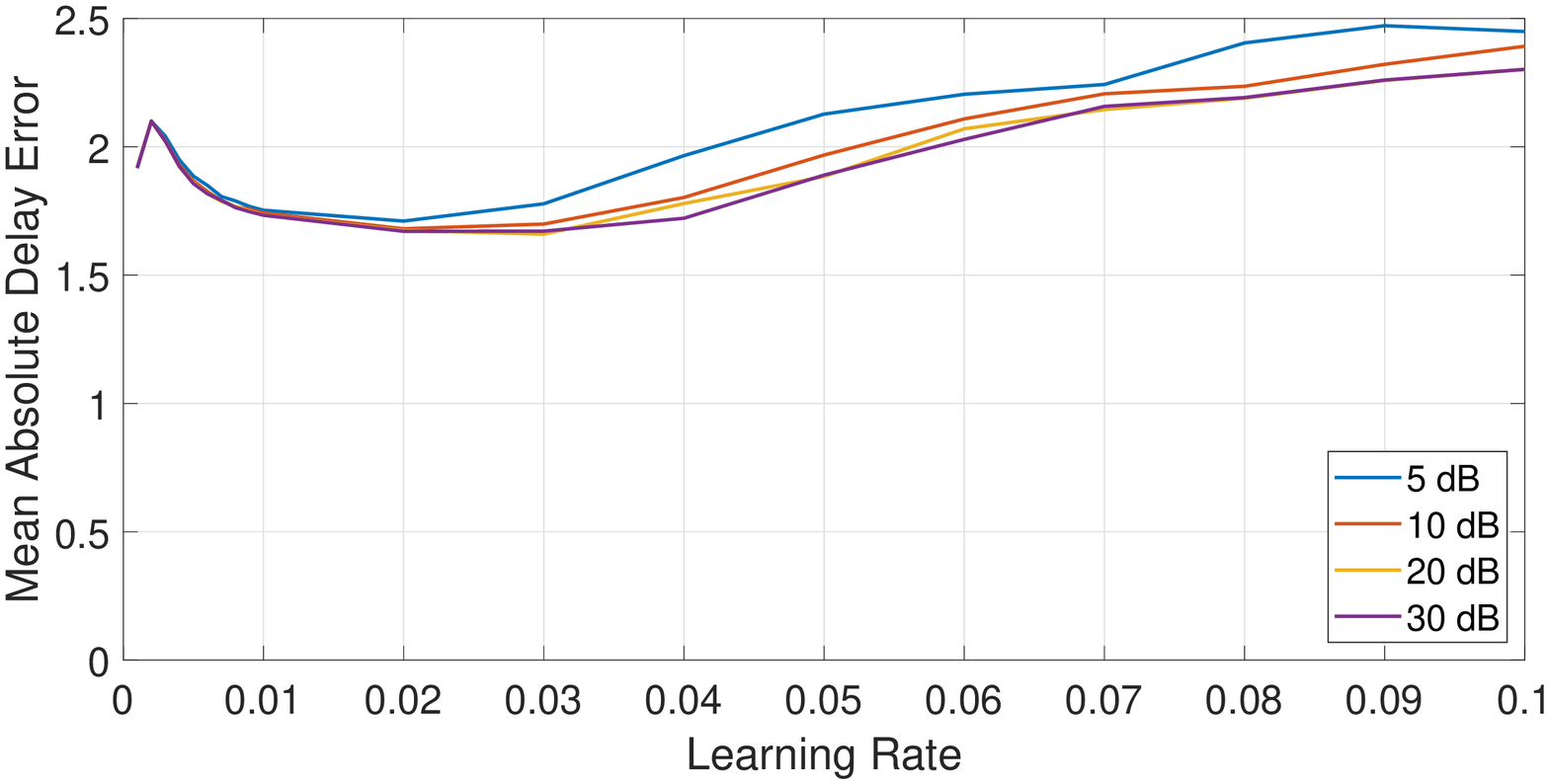}
    }
    \subfloat[]{
	    \includegraphics[width=0.32\linewidth]{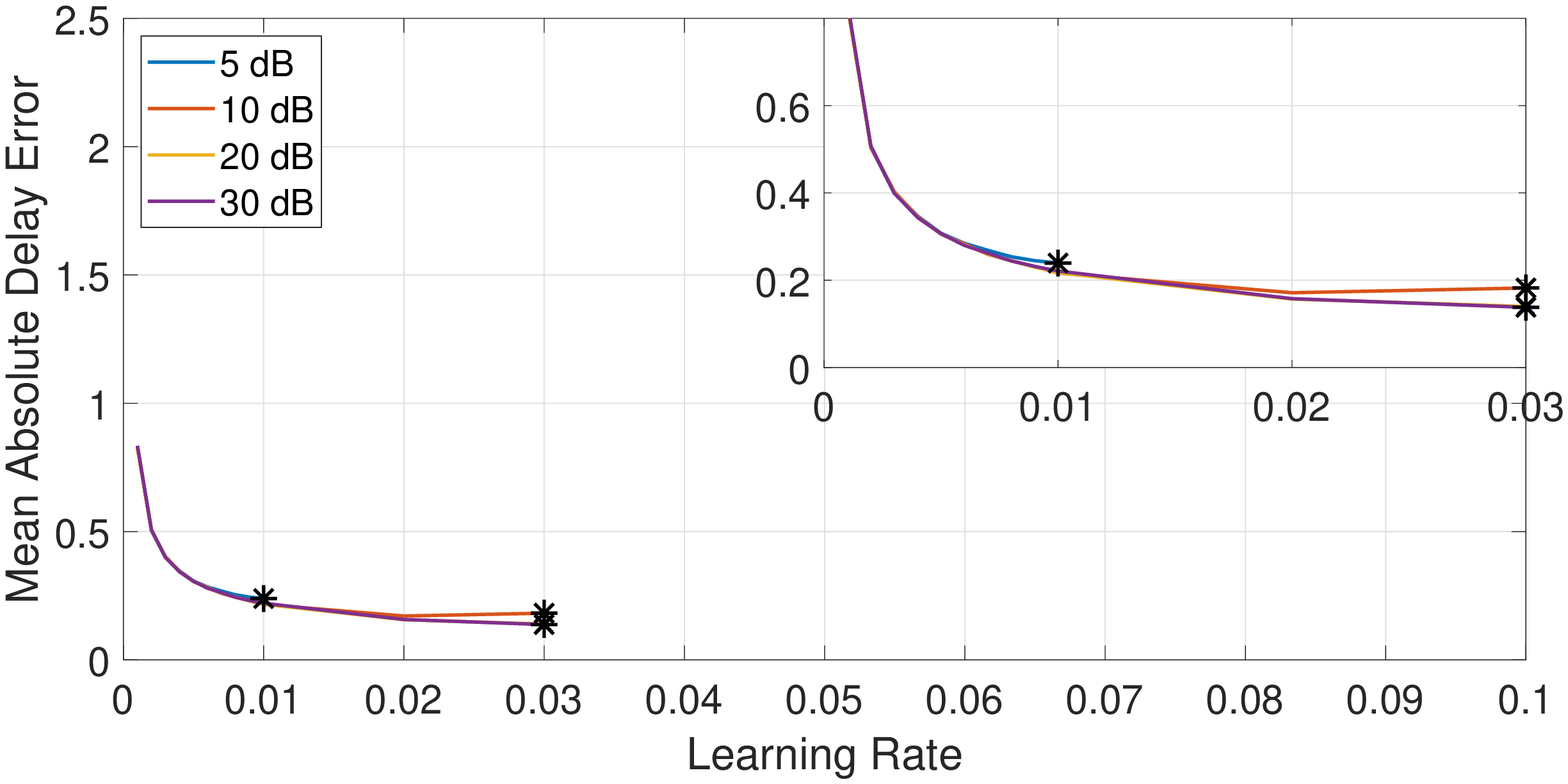}
    }
	\caption{Average mean absolute delay error for different learning rates in the large step change scenario (a) The NAAP. (b) The ETDE (c) The Sun algorithm. The marker $\mathbf{\ast}$ indicates the point at which the algorithm becomes numerically unstable.}
	\label{fig:largeStep}
\end{figure*}

Figures~\ref{fig:smallStep} and~\ref{fig:largeStep} show the evolution of the average mean absolute delay error as the learning rate increases for each of the algorithms in the small and large step change scenarios, respectively. Note the Sun algorithm becomes numerically unstable as the learning rate increases.


\printbibliography

\end{document}